\documentclass[aps,prb,twocolumn,superscriptaddress,showpacs]{revtex4-1}

\usepackage{graphicx}
\usepackage{amsmath}
\usepackage{graphics}
\usepackage{exscale}
\usepackage{epsfig}

\begin{document}

\preprint{APS/123-QED}

\title{Magnetic field induced quantum criticality and the Luttinger sum
rule}

\author{Y. Nishikawa}
\affiliation{Dept. of Physics, Osaka City University, Sumiyoshi-ku, Osaka 558-8585}
\author{O. J. Curtin}%
\author{A. C. Hewson}
\author{D. J. G. Crow}
\affiliation{Dept.of Mathematics,Imperial College,London SW7 2AZ,UK.}%


\date{\today}

\begin{abstract}
We show 
that when there is a sudden transition from a small to a large 
Fermi surface at a field
induced quantum critical point - 
similar to what may have been observed in some heavy fermion compounds -  
an additional term has to be taken into account in the Luttinger-Friedel
 sum rule.
We calculate this
additional term for a local model which has a field induced quantum 
critical point (QCP) and show that
it changes abruptly at the transition, such that it satisfies a generalized 
Luttinger-Friedel sum rule on each side of the transition, and
 characterizes the two Fermi liquid phases separated by the QCP as a
 discrete (topological) index.
\end{abstract}


\maketitle


%
\section{Introduction}
Quantum critical behavior has been induced in many heavy fermion compounds by reducing the critical temperature
of a low temperature phase transition to zero via techniques such as pressure, alloying or an applied
magnetic field
\cite{rf1,rf2}
. 
The critical behavior of some of these
compounds has been understood in the framework of the
Wilson renormalization group approach, when the quantum
mechanical fluctuations as well as the thermal fluctuations
of the order parameter are taken into account
\cite{rf3,rf4}
.
For others, however, where the fluctuations appear to be
predominantly local, the extended Wilson approach does
not satisfactorily explain the experimental results. There
is as yet no fully comprehensive theory of this type of
critical behavior so it is a very active area of research,
both experimental and theoretical 
\cite{rf5,rf6,rf7}
.
In some heavy fermion compounds,  a sudden jump from a small volume to a
large volume Fermi surface
may have occurred at the
field induced transition
\cite{rf9, rf10,rf11}
.
As the Luttinger sum rule relates the density of electrons in partially filled
bands to the volume of the Fermi surface, this change in volume indicates a sudden change in the degree of localization
of the f-electrons. This Fermi surface volume
change may also be viewed as supporting the Kondo collapse
scenario where the behavior at the quantum critical
point has been interpreted as arising from a sudden
disappearance of the states at the Fermi level associated
with a Kondo resonance or renormalized f-band
\cite{rf7,rf12}
.
To investigate this more fully, we first of all review the steps
in the derivation of Luttinger theorem to assess the implications
of a system having electrons in partially filled
bands and Fermi surfaces with different volumes. We
then relate this behavior to that of a local model with a
field induced quantum critical point which has a Kondo
regime with renormalized f-states at the Fermi level.

We consider a system of non-interacting electrons in
eigenstates specified by the set of quantum numbers denoted
by $\alpha$ 
with single particle energies $E_{\alpha}$ . 
When two-body
interaction terms are included the single-particle
Green's function $G_{\alpha}(\omega)$ can be written in the form,
\begin{equation}
G_{\alpha}(\omega)
=
\frac{1}{\omega-\epsilon_{\alpha}-\Sigma_{\alpha}(\omega)},
\end{equation}
where 
$\epsilon_{\alpha}=E_{\alpha}-\mu$
is the single particle excitation energy
relative to the chemical potential 
$\mu$ and $\Sigma_{\alpha}(\omega)$ is the
proper self-energy term. The expectation value for the
total number of electrons in the system, 
$N$, is then given by
\begin{equation}
N=-\frac{1}{\pi}\sum_{\alpha}
\int_{-\infty}^{0}
\lim_{\delta\to 0+}
\left[
{\rm Im}G_{\alpha}(\omega^{+})
\right]
d\omega,\label{Neq}
\end{equation}
where 
$\omega^{+}=\omega+i\delta \ (\delta>0)$ and we have taken
$\mu=0$. 
On replacing $G_{\alpha}(\omega^{+})$ in the integrand by
$\left(1-\Sigma^{\prime}_{\alpha}(\omega^{+})\right)G_{\alpha}(\omega^{+})+\Sigma^{\prime}_{\alpha}(\omega^{+})G_{\alpha}(\omega^{+})$
, where the prime indicates a derivative with respect to $\omega$, 
the first integral can be performed to give
\begin{equation}
N=\sum_{\alpha}
\left[
\frac{1}{2}
-
\frac{1}{\pi}
\arctan
\left(
\frac
{\epsilon_{\alpha}+\Sigma^{R}_{\alpha}(0^{+})}
{\delta+\Sigma^{I}_{\alpha}(0^{+})}
\right)
\right]_{\delta\to 0^{+}}
+
\sum_{\alpha}I_{\alpha},\label{Neqmod}
\end{equation}
where
\begin{equation}
I_{\alpha}
=
-\frac{1}{\pi}
\int_{-\infty}^{0}
{\rm Im}
\left[
\frac
{\partial\Sigma_{\alpha}(\omega^{+})}
{\partial\omega}
G_{\alpha}(\omega^{+})
\right]_{\delta\to 0^{+}}
d\omega,
\end{equation}
and  $\Sigma^{R}_{\alpha}(\omega^{+})$ and
$\Sigma^{I}_{\alpha}(\omega^{+})$ are the real and imaginary
parts of $\Sigma_{\alpha}(\omega)$ respectively.

Note that in deriving the general Eqn. (\ref{Neq}) no assumptions
of translational invariance or periodicity have been
made, so it applies to systems with impurities, including
the Friedel sum rule for the single impurity Anderson
model if it is expressed in terms of the diagonalized one electron
states. It also takes into account any partially
filled localized or atomic states.

Luttinger
\cite{rf13,rf13a} 
showed, within perturbation theory, that
the imaginary part of the self-energy at the Fermi level
vanishes, i.e. $\left[\Sigma_{\alpha}^{I}(0^{+})\right]_{\delta\to
0^{+}}=0$ due to phase space restrictions
on scattering processes. If this condition holds
and  $\Sigma_{\alpha}^{R}(0)$ is finite, which we denote as condition (i),
then we can simplify Eqn. (\ref{Neqmod}) and rewrite it in the form,
\begin{equation}
N=\sum_{\alpha}
\left[
1-\theta\left(
\epsilon_{\alpha}
+
\Sigma_{\alpha}^{R}
(0)
\right)
\right]
+
\sum_{\alpha}I_{\alpha}.\label{Neqmod1}
\end{equation}
We will refer to this equation as a generalized Luttinger-
Friedel sum rule (GLFSR).

Condition (i) is required to be able to define a Fermi
surface. If the one-electron states correspond to Bloch
states in a lattice with periodic boundary conditions so
that the index $\alpha$ denotes a wave-vector {\bf k} and band index
$n$ and a spin index $\sigma$, then if (i) holds, a Fermi surface
can be defined as the locus of points ${\bf k}_{\rm F}$ which satisfies
\begin{equation}
\epsilon_{n,{\bf k}_{\rm F}}
+
\Sigma^{R}_{n,{\bf k}_{F}}(0)
=
0.
\end{equation}
In this case Eqn. (\ref{Neqmod1}) becomes
\begin{equation}
N=
\sum_{{\bf k},\sigma}
\left[
1
-
\theta\left(
\epsilon_{{\bf k},\sigma}
+
\Sigma^{R}_{{\bf k},\sigma}(0)
\right)
\right]
+
\sum_{{\bf k},\sigma}
I_{{\bf k},\sigma}.
\end{equation}
The Luttinger sum rule relating the total number of electrons
in partially filled bands to the sum of the volumes
of the spin up and spin down Fermi surfaces follows if the
Luttinger integral, $\sum_{{\bf k},\sigma}I_{{\bf k},\sigma}=0$.

If there is a change in the volume of the Fermi surface at a transition,
yet no change in the total number
of electrons in the states contributing to the sum rule,
it follows that the term, $\sum_{{\bf k},\sigma}I_{{\bf k},\sigma}$ ,
cannot be zero
through the transition
. We will exploit the consequences of this
observation for a particular model later in this paper.

The relation in Eqn. (\ref{Neqmod1}) can be given an interpretation
in terms of quasiparticles if  
$\Sigma^{R}_{\alpha}(\omega)$
 has a finite
derivative with respect to $\omega$ at $\omega=0$ and the derivative,
$\partial\Sigma^{I}_{\alpha}(\omega)/\partial\omega$
, is zero evaluated at $\omega=0$, which we denote
as condition (ii). 
The excitation energy of a quasiparticle
$\tilde{\epsilon}_{\alpha}$ can be defined as 
$\tilde{\epsilon}_{\alpha}=z_{\alpha}\left(\epsilon_{\alpha}+\Sigma^{R}_{\alpha}(0)\right)$
, where $z_{\alpha} (<1)$
is the quasiparticle weight given by
\begin{equation}
z^{-1}_{\alpha}
=
\left[
1
-
\frac
{\partial\Sigma^{R}_{\alpha}(\omega)}
{\partial\omega}
\right]_{\omega=0}.
\end{equation}
If conditions (i) and (ii) are satisfied one can define a
total quasiparticle density of states, $\tilde{\rho}(\omega)$
 via
\begin{equation}
\tilde{\rho}(\omega)
=
\sum_{\alpha}
\delta\left(
\omega-\tilde{\epsilon}_{\alpha}
\right).
\end{equation}
Note that the quasiparticle density of states is different
from the quasiparticle contribution to the spectral density
of $\sum_{\alpha}G_{\alpha}(\omega)$. In terms of the quasiparticles Eqn.
(\ref{Neqmod1}) takes the form,
\begin{equation}
N=
\sum_{\alpha}
\left[
1
-
\theta(\tilde{\epsilon}_{\alpha})
\right]
+
\sum_{\alpha}
I_{\alpha},
\end{equation}
which sums over the number of occupied free quasiparticle
states.
Equivalently the first term of the right hand
side can be expressed as an integral over the free quasiparticle
density of states $\tilde{\rho}(\omega)$ up to the Fermi level,
\begin{equation}
N=\int_{-\infty}^{0}
\tilde{\rho}(\omega)d\omega
+
\sum_{\alpha}I_{\alpha}.
\end{equation}

The existence of long-lived quasiparticle excitations in
the neighborhood of the Fermi level is one of the basic
assumptions in the phenomenological Fermi liquid theory of
Landau, and the microscopic theory provides a criterion
for such excitations to exist. For a Fermi liquid a further
condition is usually invoked, that the one-electron excitations
of the interacting system are adiabatically connected
to those of the non-interacting system, which we
denote as condition (iii). The proof of the Luttinger sum
rule, that $\sum_{\alpha}I_{\alpha}=0$, as given by Luttinger and Ward
\cite{rf14}
requires this third condition (iii) to hold, and adiabatic
continuity would appear to be a general requirement in
an alternative non-perturbative derivation for a Fermi
liquid
\cite{rf15,rf15a}
. However, it is possible for quasiparticle excitations
to exist at the Fermi level in a phase in which condition
(iii) does not hold, and we have given an example
in earlier work
\cite{rf15,rf15a}
. Here we consider a local system which,
though not directly applicable as a model of heavy fermion compounds,
has a field induced quantum critical point and has a sum
rule that mirrors the change in volume of the
Fermi surface as may have been observed in some heavy fermion compounds.
\begin{figure}[b]
\includegraphics[width=0.4\textwidth]{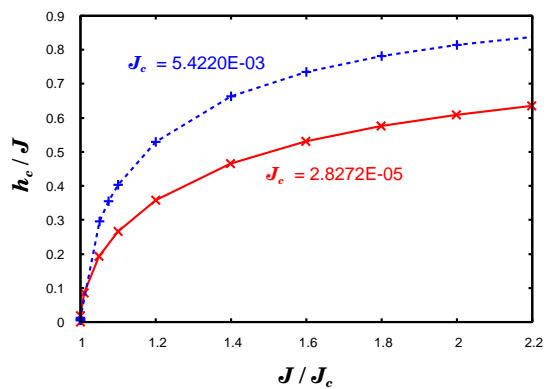}
\caption{ (Color online) A plot of $h_{c}/J$, where $h_{c}$ is critical field
to induce a transition, as a function of the interaction parameter
$J$ for a particle-hole symmetric model with $U/\pi\Delta  = 5, J_{c} = 2.8272\times 10^{-5}$ and for an asymmetric model with
$\epsilon_{f}/\pi\Delta = -0.659, U/\pi\Delta  = 0.5, J_{c} =
 5.4220\times 10^{-3}$(dashed line).}
\end{figure}

\section{Model}
The model we consider is a version of the two impurity
Anderson model in the presence of magnetic field.
The Hamiltonian for this model takes the form, 
${\mathcal H}=\sum_{i=1,2}{\mathcal H}_{i}+{\mathcal H}_{12}$
, where ${\mathcal H}_{i}$ corresponds to an individual
Anderson impurity model with channel index $i$,
\begin{eqnarray}
{\mathcal H}_{i}
&=&
\sum_{\sigma}
\epsilon_{f,i,\sigma}
f^{\dagger}_{i,\sigma}f_{i,\sigma}
+
\sum_{k,\sigma}
\epsilon_{k,i}
c^{\dagger}_{k,i,\sigma}c_{k,i,\sigma}\nonumber\\
& &\mbox{}
+
\sum_{k,\sigma}
\left(
V_{k,i}f^{\dagger}_{i,\sigma}c_{k,i,\sigma}+{\rm h.c.}
\right)
+
U_{i}n_{f,i,\uparrow}n_{f,i,\downarrow},
\end{eqnarray}
where $f^{\dagger}_{i,\sigma}$, $f_{i,\sigma}$, are creation and annihilation operators
for an electron at the impurity site in channel $i$ and spin
component $\sigma=\uparrow, \ \downarrow$,  
and energy level 
$\epsilon_{f,i,\sigma}=\epsilon_{f,i}-h\sigma$,
where $h = g\mu_{\rm B}H/2$, $H$ is a local magnetic field, $g$ is the
g-factor and $\mu_{\rm B}$ the Bohr magneton. The creation and
annihilation operators 
$c^{\dagger}_{k,i,\sigma}$, $c_{k,i,\sigma}$
 are for partial wave
conduction electrons with energy $\epsilon_{k,i}$ in channel $i$, each
with a bandwidth $2D$ with $D = 1$. With the assumption
of a flat wide conduction band the hybridization factor,
$\Delta_{i}(\omega)=\pi\sum_{k}|V_{k,i}|^{2}\delta(\omega-\epsilon_{k,i})$, can be taken as constant
independent of $\omega$.
For the inter-impurity interaction Hamiltonian ${\mathcal H}_{12}$ we
take into account an antiferromagnetic exchange term $J$
and a direct interaction $U_{12}$,
\begin{equation}
{\mathcal H}_{12}
=
2J{\bf S}_{f,1}\cdot{\bf S}_{f,2}
+
U_{12}
\sum_{\sigma,\sigma^{\prime}}
n_{f,1,\sigma}n_{f,2,\sigma^{\prime}}.
\end{equation}

\begin{figure}[b]
\includegraphics[width=0.4\textwidth]{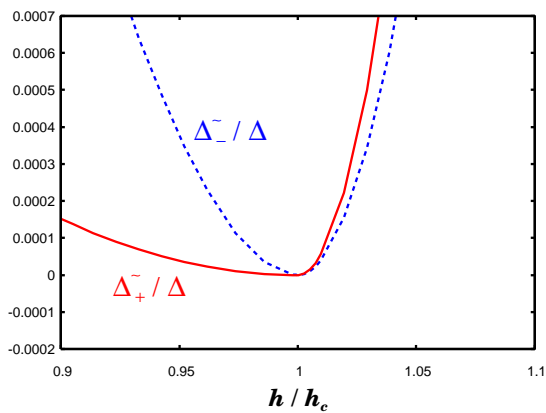}
\caption{(Color online) A plot of the quasiparticle weight factors, 
$z_{\uparrow} = \tilde{\Delta}_{\uparrow}/\Delta$ and $z_{\downarrow} =\tilde{\Delta}_{\downarrow}/\Delta$  
as a function of the magnetic field $h/h_{c}$ in the region of the transition for the
parameter set 
$\epsilon_{f}/\pi\Delta  = -0.659, U/\pi\Delta  = 0.5, J_{c} = 5.4420\times 10^{-3}, J = 5.8482\times10^{-3}$, where $h_{c}/\pi\Delta  = 0.20795159$.}
\end{figure}

The Kondo version of the two impurity model has a
long history and was originally put forward to study the
competition between the Kondo effect and the intersite
RKKY interaction in heavy fermion systems. It was in
this context that it was shown to have $T  = 0$ phase
transition
\cite{rf16,rf16a,rf16b,rf16c,rf16d,rf16e}
. Studies based on the two impurity Anderson
model in the absence of a magnetic field have revealed
a discontinuous change in spectral density at the
Fermi level at the transition
\cite{rf17,rf17a}
. Other studies have shown
that the transition is very robust, occurring away from
particle-hole symmetry
\cite{rf18} 
 and even away from the Kondo
regime with $U_{1} = U_{2} = U_{12} = 0$
\cite{rf15,rf15a}
\cite{rf19,rf19a}
. The basic picture
that emerges is that as the inter-site interaction $J$ is increased
from $J = 0$, where the sites are Kondo screened
in the case of large $U$, a new universal low energy scale
$T^{\ast}$
is induced such that on the approach to the transition
at $J = J_{c}$, $T^{\ast}\to 0$. For $J > J_{c}$, the screening is
then predominantly due to the existence of an induced
local singlet state, which we will refer to as a localized
dimer singlet (LDS) state to distinguish it from the singlet
state associated with the Kondo resonance in the
regime $J < J_{c}$. The LDS state can equally well be interpreted
as a resonant valence bond (RVB) state.

To demonstrate the field induced transition we consider
the model with identical impurities and baths, so
we drop the index $i$, in the presence of a local magnetic
field $H$. 
We consider a case first of all with particle-hole
symmetry and parameters $U/\pi\Delta=5$ ($\epsilon_{f}=-U/2$) and
$\pi\Delta  = 0.01$ (this value will be taken in all calculations)
where $J_{c}$ is $2.8272\times10^{-5}$. 
In the absence of a magnetic
field and $J < J_{c}$, 
there is a Kondo resonance at the Fermi
level with a spectral density 
$\rho_{f}(0)=1/\pi\Delta$. 
In the LDS phase for $J > J_{c}$, however, the Kondo resonance disappears
leaving a pseudogap with $\rho_{f}(0)=0$, reflecting the
fact that it costs a finite energy to remove an f-electron
and break apart the LDS state (this can be interpreted as
the binding energy of the LDS state). 
This situation can be reversed if we start in the phase with $J > J_{c}$ and introduce
a magnetic field, as an increase in the field leads
to a decrease in the energy gain from the LDS formation.
At a critical value $h = h_{c}$, there is an unstable non-Fermi
liquid fixed point, such that for $h > h_{c}$ the system reverts
to a phase with the states at the Fermi level associated
with the Kondo resonances fully restored, but now polarized
by the magnetic field. The greater the value of
$J/J_{c}$, the larger field field required to induce this transition
as can be seen in Fig. 1 where we give a plot of
$h_{c}/J$ as a function of $J/J_{c}$. As similar transition occurs
in the case we consider away from particle-hole symmetry
with parameters, 
$\epsilon_{f}/\pi\Delta  = -0.659, U/\pi\Delta  = 0.5, J_{c} = 5.4220\times 10^{-3}$, and the critical field $h_{c}$ for
this case is also shown in Fig. 1. The only qualitative
difference with the particle-hole symmetric case is that
for $J > J_{c}$, in the absence of a magnetic field, there is
some residual f-spectral weight at the Fermi level, as we
will show later.

\begin{figure}[b]
\includegraphics[width=0.4\textwidth]{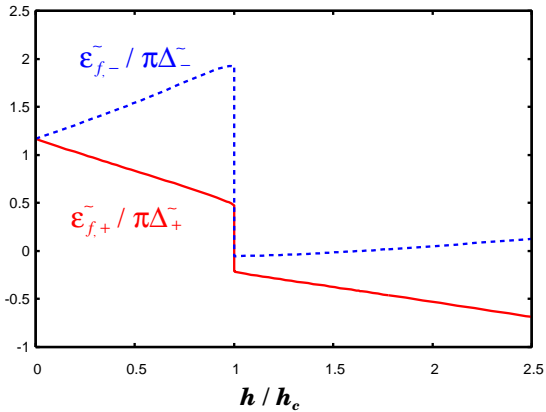}
\caption{(Color online) A plot of the ratios, 
$\tilde{\epsilon}_{f,+}/\tilde{\Delta}_{+} \ (\tilde{\epsilon}_{f,\uparrow}/\tilde{\Delta}_{\uparrow})$
and 
$\tilde{\epsilon}_{f,-}/\tilde{\Delta}_{-} \ (\tilde{\epsilon}_{f,\downarrow}/\tilde{\Delta}_{\downarrow})$
, as a function of the magnetic field
$h/h_{c}$ for parameter set given in Fig. 2.}
\end{figure}

\section{Results, Discussions and Conclusions}
To understand this transition in more detail, we calculate
the renormalized parameters characterizing the low
energy excitations as a function of the magnetic field. We
can then see how these parameters vary with $h$ and follow
their behavior as we cross the transition. We define
a renormalized energy level 
$\tilde{\epsilon}_{f,\sigma}$
,  and resonance width 
$\tilde{\Delta}_{\sigma}$
via
\begin{equation}
\tilde{\epsilon}_{f,\sigma}
=
z_{\sigma}
\left(
\epsilon_{f,\sigma}
+
\Sigma_{\sigma}(0)
\right), \ \ 
\tilde{\Delta}_{\sigma}
=
z_{\sigma}\Delta_{\sigma},\label{tilde-e}
\end{equation}
where $\Sigma_{\sigma}(\omega)$ is the self-energy for an individual impurity
zero temperature causal Green's function, 
$G_{f,\sigma}(\omega)$, given by 
\begin{figure}[h]
\includegraphics[width=0.4\textwidth]{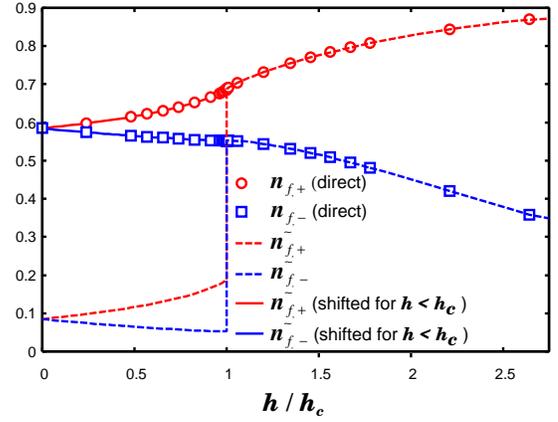}
\caption{(Color online) A plot of the spin up and spin down
local occupation values 
$n_{f,+}$, and $n_{f,-}$ 
from a direct calculation as a function of the magnetic field $h$ for the
parameter set of Fig. 2. 
Also shown are the values of 
$\tilde{n}_{f,+}$
and 
$\tilde{n}_{f,-}$ and the corresponding values shifted by $+\frac{1}{2}$ for $h < h_{c}$.}
\end{figure}

\begin{figure}[h]
\includegraphics[width=0.4\textwidth]{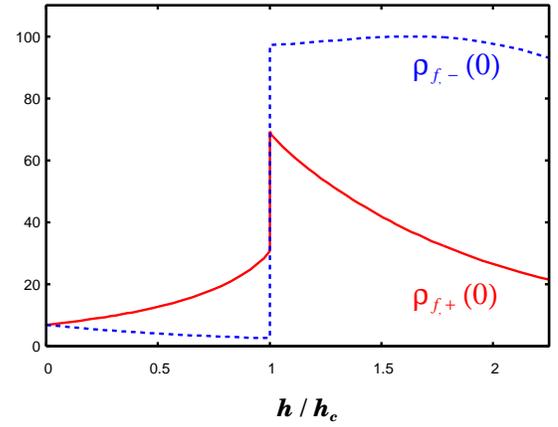}
\caption{(Color online) A plot of the spin up and spin down
local spectral densities, $\rho_{f,+}$ and $\rho_{f,-}$, as a function of the
magnetic field $h/h_{c}$ for the asymmetric parameter set of Fig.
2, showing the sudden increase at the quantum critical point
as $h$ increases through $h_{c}$.}
\end{figure}

\begin{figure}[h]
\includegraphics[width=0.4\textwidth]{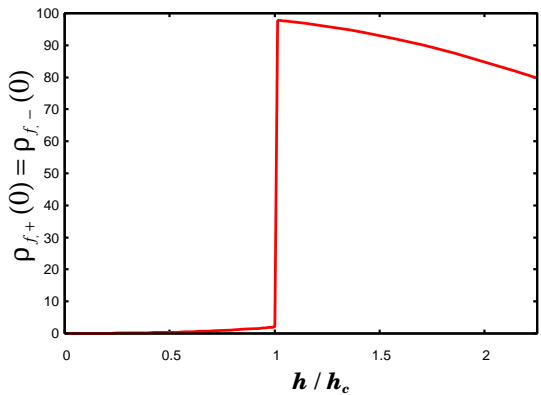}
\caption{(Color online) A plot of the spin up and spin down
local spectral densities, $\rho_{f,+}$ and $\rho_{f,-}$ ( $\rho_{f,+}$ =
 $\rho_{f,-}$ ), as a function of the
magnetic field $h/h_{c}$ for the particle-hole symmetric parameter
set, $U/\pi\Delta  = 5, J = 3.0354436\times 10^{-5}, J_{c} = 2.8272\times 10^{-5}$
showing the sudden increase at the quantum critical point as
$h$ increases through $h_{c}$.}
\end{figure}
\begin{equation}
G_{f,\sigma}(\omega)
=
\frac{1}
{
\omega-\epsilon_{f,\sigma}+i\Delta{\rm sgn}(\omega)
-
\Sigma_{\sigma}(\omega)
},
\end{equation}
and
$z_{\sigma}=\left(1-\partial\Sigma_{\sigma}(\omega)/\partial\omega\right)^{-1}_{\omega=0}$. 

These renormalized parameters can be deduced from NRG excitations in the
neighborhood of the low-energy fixed point
\cite{rf20}
. 
The parameter that reflects the loss of quasiparticles as the transition
is approached, is the quasiparticle weight factor
$z_{\sigma}=\tilde{\Delta}_{\sigma}/\Delta$ .
 This is plotted in Fig. 2 for both spin
types as a function of $h/h_{c}$ for a particle-hole asymmetric
model with the parameter set, 
$\epsilon_{f}/\pi\Delta= -0.659, U/\pi\Delta  = 0.5, J_{c} = 5.4220\times10^{-3}, J = 5.8482\times 10^{-3}$,
where $h_{c}/\pi\Delta  = 0.20795159$, showing that the quasiparticles
disappear as the non-Fermi liquid fixed point is
approached $h\to h_{c}$. 
Though all the quasiparticle parameters
tend to zero as $h\to h_{c}$, their ratios remain finite.
The ratio 
$\tilde{\epsilon}_{f,\sigma}/\pi\tilde{\Delta}_{\sigma}$
, however, has different limiting
values on the two sides of the transition as can be seen
in the plot in Fig. 3 for the same parameter set, and
corresponds to a phase shift of $\pi/2$. 
The discontinuity
in 
$\tilde{\epsilon}_{f,\sigma}/\pi\tilde{\Delta}_{\sigma}$
arises from a discontinuity in the self-energy
term $\Sigma_{\sigma}(0)$ in Eqn. (\ref{tilde-e}). It has also been confirmed from
a direct NRG calculation of $\Sigma_{\sigma}(\omega)$. This discontinuity
has consequences for the sum rule giving the occupation
numbers for the impurity levels.

The generalized Luttinger-Friedel sum rule for each
spin component, in terms of renormalized parameters,
takes the form,
\begin{equation}
n_{f,\sigma}
=
\frac{1}{2}
-
\frac{1}{\pi}
\arctan
\left(
\frac
{\tilde{\epsilon}_{f,\sigma}}
{\tilde{\Delta}_{\sigma}}
\right)
+
I_{f,\sigma},
\end{equation}
where $n_{f,\sigma}$,  is occupation number for the local level with
spin $\sigma$ on each impurity.
We denote the value of this
expression with $I_{f,\sigma} = 0$ by 
$\tilde{n}_{f,\sigma}$ 
.
 The value of $n_{f,\sigma}$, can be calculated from a direct NRG calculation or indirectly
from an integration over the spectral density 
$\rho_{f,\sigma}(\omega)$
of
the Green's function 
$G_{f,\sigma}(\omega)$, 
and 
$\tilde{n}_{f,\sigma}$
from the renormalized parameters. 
Results for $n_{f,\sigma}$,  as a function of $h$ is
shown in Fig. 4 together with the corresponding results
for $\tilde{n}_{f,\sigma}$ .
 For $h > h_{c}$ there is complete agreement between
$n_{f,\sigma}$,  and $\tilde{n}_{f,\sigma}$
,  as expected from the standard Friedel sum
rule. 
For $h < h_{c}$ we find $I_{f,\sigma}=\frac{1}{2}$ so the shifted value,
$\tilde{n}_{f,\sigma}+\frac{1}{2}$ , is compared with 
$\tilde{n}_{f,\sigma}$
in this range. What is
clear is that the particle occupation number is continuous
through the transition, whereas 
$\tilde{n}_{f,\sigma}$
is not. 
Note that in the situation with 
$\epsilon_{f}+U/2 > 0$
, $I_{f,\sigma}$,  has the opposite
sign if the arctan is evaluated over the same branch. 
As
$\tilde{n}_{f,\sigma}=n_{f,\sigma}-\frac{1}{2}$
in the range with a LDS, we can interpret
$\tilde{n}_{f,\sigma}$
as the contribution from the residual quasiparticles
that remain for $h < h_{c}$. 
Further evidence backing this interpretation can be seen by looking at the value of the
local spectral density at the Fermi level, 
$\rho_{f,\sigma}(0)$
, which is given by
\begin{equation}
\rho_{f,\sigma}(0)
=
\frac{1}{\pi\Delta}
\frac{\tilde{\Delta}^{2}_{\sigma}}
{\tilde{\epsilon}^{2}_{f,\sigma}+\tilde{\Delta}^{2}_{\sigma}},
\end{equation}
in terms of renormalized parameters.
The values of
$\rho_{f,\sigma}(0)$
for the up and down spins,
$\sigma = +(\uparrow), \ -(\downarrow)$
, are
shown in Fig. 5 as a function of $h$ through the transition.
As the field value $h$ is reduced from above $h_{c}$ to below $h_{c}$
there is a sudden loss of spectral weight at the transition.
This is just what is expected as a LDS is formed, localizing
$\frac{1}{2}$ of an electron of each spin type. As this does
not account for all the f-electrons, there are still some
non-localized quasiparticle states at the Fermi level corresponding
to the residual f-electrons. This Fermi liquid
state ($h < h_{c}$), which has well defined quasiparticles but
does not satisfy the usual sum rule due to the Luttinger
integral contribution, corresponds to what in a Kondo
lattice model has been described as a fractionalized Fermi
liquid
\cite{rf21}
. It is also similar to the phenomenological duality
model proposed for some heavy fermion systems
\cite{rf22}
 as
well as for the interpretation of the magnetism in some
3d transition metals
\cite{rf23,rf23a}
. 
The transition as the field $h$ is
increased through $h_{c}$ can therefore be interpreted as a
transition from a fractionalized to a normal Fermi liquid.
Recently a fermionic and bosonic dimer model for a fractionalized
Fermi liquid has been put forward to describe
the pseudogap phase of cuprate superconductors
\cite{rf24,rf24a}
 and
to explain the observed quantum oscillations
\cite{rf25}
.

It is interesting to examine the case at particle-hole
symmetry where for $J > J_{c}$ and $h = 0$, there are no
quasiparticle states and the local spectral density at the
Fermi level is zero. In Fig. 6 the results for 
$\rho_{f,\sigma}(0)$
for
the case with 
$U/\pi\Delta  = 5, J = 3.0354436\times10^{-5}$ and
$J_{c} = 2.8272\times10^{-5}$
, are shown across the transition.
Though 
$\rho_{f,\sigma}(0)=0$
for $h = 0$, for finite $h < h_{c}$ there
are some non-localized quasiparticle states induced giving
a small but finite values of 
$\rho_{f,\sigma}(0)$
in this regime. 
The quasiparticle contribution, however, only becomes significant
with the sudden jump in 
$\rho_{f,\sigma}(0)$
at the transition
which then slowly decreases with increasing $h$ as the spin
up and spin down Kondo resonances move away from the
Fermi level.

There is a 
 correspondence of these results
with 
what may have occurred in some heavy fermion compounds
\cite{rf10,rf11}
at the
field induced transition
: the sudden change
in the number of quasiparticle states at the Fermi level
and the corresponding increase in the volume of the Fermi
surface in line with that expected from the generalized
Luttinger-Friedel sum rule. 
( There is a difference in that
the phase with $h < h_{c}$ 
 has features
which have led it being identified as a non-Fermi
liquid
\cite{rf9}
. )
While in the impurity case it is a Fermi liquid
for finite $h$ and $T = 0$, it does have finite temperature
non-Fermi liquid terms in the vicinity of fixed point at
$h = h_{c}$. There are as yet no exact results for a lattice
model of these materials. However, a recent study based
on a large-N mean field Hamiltonian does indicate a suppression
by the magnetic field of the RKKY coupling relative
to the Kondo coupling leading to a transition from
a fractionalized Fermi liquid to a heavy Fermi liquid with
Kondo correlations
\cite{rf26}
. If one can understand what happens
to the Luttinger sum rule in passing through a field
dependent quantum critical point in the model we have
studied, then this can provide a guide to what can happen
in more complicated models.

\section{Acknowledgement}
One of us (Y.N.) acknowledges the support by JSPS
KAKENHI Grant No.JP15K05181.

\bibliography{GLTOP, artikel, biblio1}

\end{document}